\documentclass[aps,preprint,superscriptaddress,amssymb,axodraw]{revtex4}

\usepackage{color}
\usepackage{graphicx}
\usepackage{epstopdf}
\usepackage{subfigure}
\usepackage{axodraw}

\newcommand\lsim{\mathrel{\rlap{\lower4pt\hbox{\hskip1pt$\sim$}}
    \raise1pt\hbox{$<$}}}
\newcommand\gsim{\mathrel{\rlap{\lower4pt\hbox{\hskip1pt$\sim$}}
    \raise1pt\hbox{$>$}}}
\def\bea{\begin{eqnarray}}
\def\eea{\end{eqnarray}}
\def\ba{\begin{array}}
\def\ea{\end{array}}
\def\bc{\begin{center}}
\def\ec{\end{center}}
\def\nn{\nonumber}

\begin{document}

\title{Higgs Phenomenology of Scalar Sequestering}

\author{Hyung Do Kim$^{1,2}$ and Ji-Hun Kim}

\affiliation{FPRD and Department of Physics, Seoul National University, Seoul, 151-747, Korea \\
$^2$ School of Natural Sciences, Institute for Advanced Study, Princeton, NJ 08540, U.S.A.}

\date{\today}

\begin{abstract}

The light Higgs mass in the MSSM is highly constrained and is predicted to be close to $M_Z$
which causes a tension between the LEP II Higgs mass bound 114 GeV and the natural electroweak symmetry breaking
in the MSSM. The usual way to increase the light CP even Higgs mass was to increase the quartic coupling of the up type Higgs.
We point out that the light Higgs mass can be increased by reducing the off-diagonal term in the mass matrix when $\tan \beta$ is moderate,
which is about 5 to 10.
As a result no mixing or a Higgs mixing angle of the opposite sign arises
and the branching ratio of Higgs decay is drastically changed.
This is possible in scalar sequestering scenario in which $\mu$ parameter can be large independently of the electroweak symmetry breaking.
We also discuss the same effect in the BMSSM.

\end{abstract}

\maketitle

\section{Introduction}

The fact that Higgs has not been discovered yet brings a question on the naturalness of weak scale supersymmetry.
The minimal supersymmetric standard model (MSSM) predicts the light Higgs to be lighter than the Z boson at tree level
and to increase it above the current experimental bound 114 GeV \cite{Barate:2003sz} requires a large one loop correction implying that the top squark (stop) is very heavy
\cite{Okada:1990vk} \cite{Haber:1990aw} \cite{Ellis:1990nz}.
One way out was to introduce a new quartic coupling in addition to one in the MSSM which goes beyond the MSSM.
In this paper we do not discuss many possibilities based on the extension of the MSSM.
Instead we ask what is the best way to reduce fine tuning in the MSSM.
It has been pointed out that large soft trilinear coupling of Higgs and stops can provide a correction
which can raise the light Higgs mass above the LEP II bound \cite{Casas:1994us} \cite{Carena:1995bx} \cite{Carena:1995wu}.
There are recent reviews on the topic in \cite{Djouadi:2005gj} \cite{Carena:2002es}.
Heavy stop can increase Higgs mass by its log correction at the expense of severe fine tuning.
Sizable correction to the Higgs mass is also possible by finite threshold correction from large trilinear coupling $A_t$ (maximal stop mixing) even for moderate value of stop mass.
In reality it is hard to obtain such a large stop mixing from high energy theory, e.g., starting from the GUT scale boundary condition.
However, it is possible if stop mass squared is negative at high energy and universal relations in sfermions and/or gauginos are given up \cite{Dermisek:2006ey}.
The least fine tuned parameter space indicates that we might live in a meta-stable vacuum.
This model independent observation has been explicitly realized in the gauge messenger models \cite{Dermisek:2006qj}.
If X and Y gauge bosons and gauginos can serve as a messenger of supersymmetry breaking,
the threshold correction at the GUT scale provides negative squark mass squared and nonuniversal gaugino masses such that
viable phenomenological spectrum can be obtained at the weak scale and at the same time the fine tuning for the electroweak symmetry breaking
can be reduced.

The characteristic feature of the large stop mixing scenario \cite{Dermisek:2006ey} is the presence of the transition scale $M_*$
at which the stop mass squared changes sign. In order to have a longevity of our universe, the scale $M_*$ should be higher than 10 TeV \cite{Riotto:1995am}.
The prediction on the stop mass squared at the GUT scale is based on the assumption that big desert exists between the GUT scale and the weak scale
such that renormalization group running is obtained only with the MSSM particles.
The trajectory can be modified at high energy with the presence of new degrees of freedom.

Recently it has been pointed out that the effect of the hidden sector running can affect the overall size of the soft parameters and maximal stop mixing can be obtained by this effect \cite{Dine:2004dv}, \cite{Cohen:2006qc}.
Especially when the hidden sector couples strongly, the effect of the hidden sector running can suppress soft scalar mass squared
and can be a solution of notorious $\mu$ and $B\mu$ problem in gauge mediation \cite{Roy:2007nz}, \cite{Murayama:2007ge}.
The effect of scalar sequestering also explains the electroweak symmetry breaking in a rather natural way even in the presence of large $\mu$
as long as $\mu$ is generated from generalized Giudice-Masiero mechanism (from Kahler potential) \cite{Perez:2008ng}.
In the simplest version of $\mu$ generation,
a direct coupling of Higgs with messengers can naturally provide $\mu$ of the order of the other supersymmetry breaking parameters.
It is the presence of $B\mu$ which is too large. There are technical solutions with new singlets in \cite{Dvali:1996cu},
\cite{Giudice:2007ca}. These singlets are heavy (or much heavier) compared to the weak scale and the weak scale physics would be exactly
that of the MSSM.

Though there is no explicit setup in which the anomalous dimensions with the desired properties are computed,
it would be worth exploring the implication of this possibility with the hope that it can be realized eventually.
The immediate outcome of the setup is the presence of the scale $M_*$ at which the scalar mass squared are largely suppressed.
It is the scale at which the strong hidden sector CFT ceases to contribute to the running and $M_*$ sets the effective scale
at which all the boundary conditions are given. Note the similarity of $M_*$ in two different scenarios.
In one scenario, the usual extrapolation of the RG running drives the stop mass squared to be negative above $M_*$.
In scalar sequestering, above $M_*$, the scalar mass squared are effectively zero as it is exponentially suppressed by large anomalous dimension
of the hidden CFT.

The second observation is in the Higgs sector. Even when $\mu$ is large (a few TeV),
$\mu^2 + m_{H_u}^2$ can be one loop suppressed compared to $\mu^2$ such that the electroweak scale can be lower than $\mu$
without having a serious fine tuning. This provides heavy gauginos and higgsinos (a few TeV) and light squarks, sleptons
and all the Higgs fields ($h^0,A,H^0,H^{\pm}$) at a few hundreds GeV. Below TeV, the MSSM spectrum would consist of sfermions
and Higgs fields only. Indeed this pattern of sparticle spectrum provides the least fine tuned electroweak symmetry breaking
as $\mu^2+m_{H_u}^2$ can be naturally light compared to gauginos and higgsinos
\footnote{Universal gaugino mass can not improve the naturalness since 1 TeV bino (factor 10 larger than sleptons of 100 GeV)
implies 6 TeV gluino which is too heavy not to cause any trouble in the Higgs mass parameters through the running
even for a short range from $M_*$ to TeV.}.
These patterns of sparticle spectrum have not been considered so far.
It would be interesting to see what would be the phenomenological signatures of this scenario.

We address this question by looking at the threshold correction to the Higgs mass when $\mu$ is large.
Large $\mu$ and also large $A_t$ can provide effective operators $H_u H_d (H_u^* H_u)$ and alters the Higgs phenomenology drastically.
The correction appears in the off-diagonal entry of the Higss mass matrix ($H_u H_d$) and also in the diagonal entry ($H_u^* H_u$).
These two corrections can increase the light Higgs mass. We emphasize the correction coming from off-diagonal element in this paper.
The branching ratio of the Higgs decay can also be seriously modified as the correction in the presence of large $\mu$
can easily cancel the original off-diagonal entry and no mixing between two scalar Higgs fields is realized even for $\tan \beta \sim 10$.

No Higgs mixing has been considered previously in the context of large $\tan \beta$ ($\tan \beta \sim 50$)
\cite{Kane:1995ek}, \cite{Baer:1998cm}, \cite{Loinaz:1998ph}.
The original off-diagonal entry is very small if $\tan \beta$ is large as it is suppressed by $\tan \beta$.
Thus even small corrections can alter the mixing angle in a drastic way and changes the Higgs phenomenology.
The difference here is that there is also an impact on the light Higgs mass when $\tan \beta$ is moderate, 5 to 10
as the correction itself is large enough. The tree level off-diagonal element is too small for large $\tan \beta$ and the net effect of off-diagonal cancellation
to the Higgs mass is negligibly small there. 
The Higgs mass is increased as a result of cancelling the off-diagonal element by the threshold correction.
For moderate $\tan \beta$, this effect on the Higgs mass can be sizable.
In the special limit of zero mixing angle between two CP even Higgs bosons,
the light CP even Higgs is just $H_u$ and the heavy CP even Higgs is $H_d$.
Then the CP even Higgs does not couple to bottom quark and tau lepton at tree level of Yukawa couplings
and can affect the Higgs branching ratio dramatically.
In the MSSM at tree level, the mixing between two CP even Higgs bosons are suppressed as $\tan \beta$ increases.
However, at the same time, the Yukawa coupling of down type quarks and leptons are enhanced by $\tan \beta$
and as a consequence the coupling of the light CP even Higgs to bottom quark and tau lepton is not suppressed.

We use DRED scheme for the computation of one loop correction
to the Higgs mass and also use the same scheme in FeynHiggs \cite{Heinemeyer:1998yj} for the plots presented here.
The parameters $\mu$ and $A_t$ are chosen to be real in this paper.

\section{Scalar Sequestering}

\begin{figure}[thb]
\subfigure[Stop mass]{
\includegraphics[width=3.in]{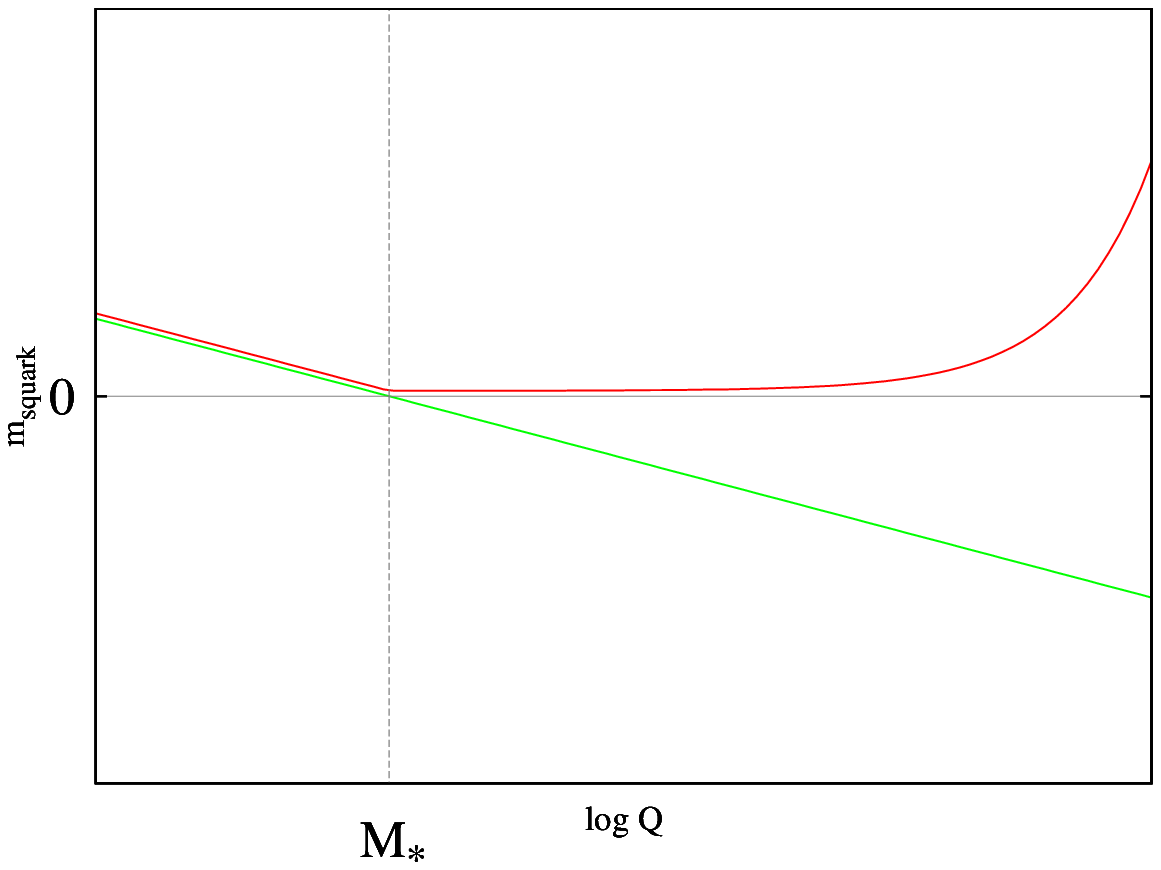}}
\subfigure[Higgs mass]{
\includegraphics[width=3.in]{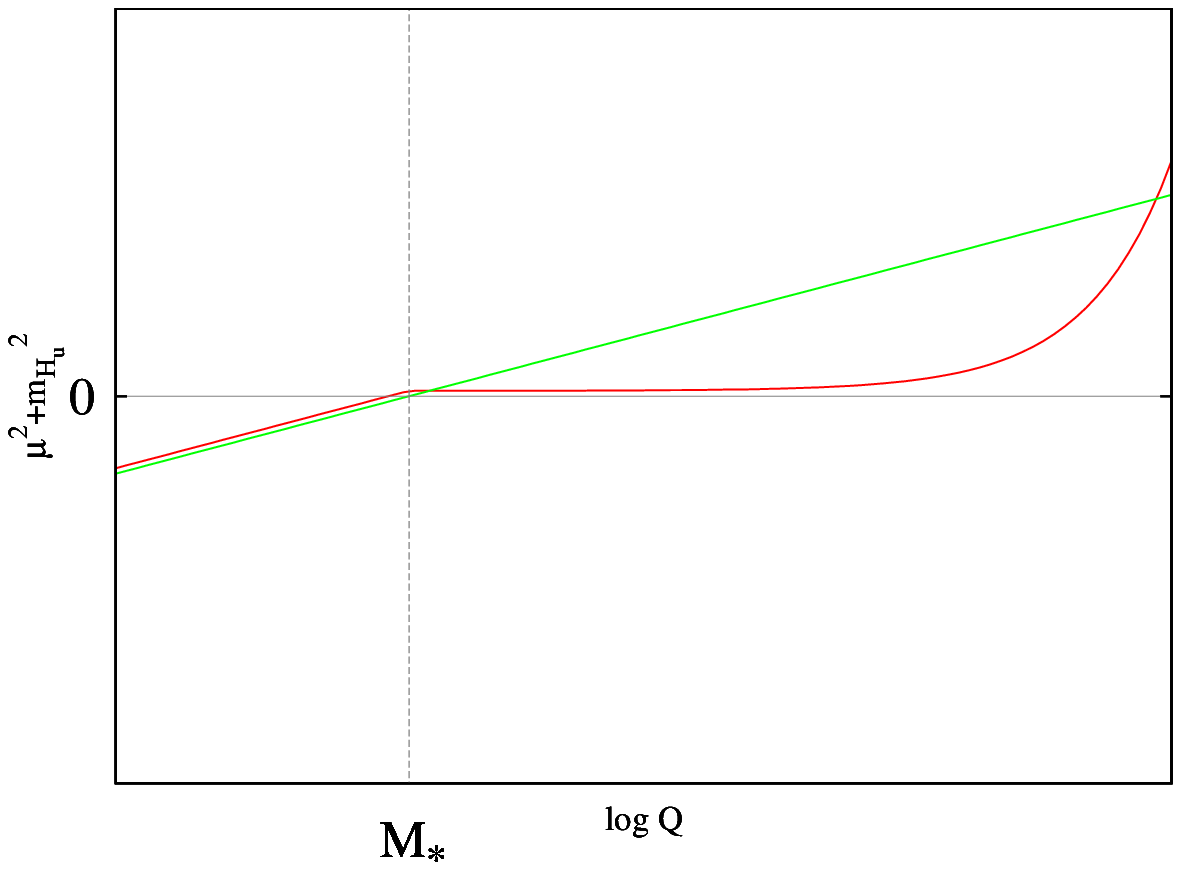}}
\caption{Running of soft parameters in tachyonic boundary condition and in scalar sequestering}
\label{fig:sequestering}
\end{figure}

In general it is very difficult to obtain small scalar mass compared to the fermion (gaugino) mass
as there is no symmetry which can forbid the scalar mass while allowing the gaugino mass.
Recently it has been pointed out that
strong hidden sector running can suppress the soft scalar mass compared to the gaugino mass
through the running from the messenger scale to the scale at which the hidden sector is integrated out.
The dimension of the operator ${\cal O}$ and ${\cal O^\dagger O}$ can be given by
\bea
\left[ {\cal O} \right] & = & d_O + \gamma,  \\
\left[ {\cal O^\dagger O} \right] & = & 2d_O + \Delta, 
\eea
where $d_O$ is the classical dimension of the operator ${\cal O}$. $\gamma$ and $\Delta$ are anomalous dimensions of the corresponding operators.
If $\Delta \neq 2 \gamma$, the hidden sector running provides a different suppression effect on ${\cal O}$ and ${\cal O^\dagger O}$.
As a result we can obtain a small soft scalar mass from the running.

Let $M_{\rm mess}$ be the scale of messengers at which the operators involving the visible sector and the supersymmetry breaking field are generated.
$M_*$ is the scale at which the strongly interacting hidden sector deviates from the conformal regime.
In between $M_{\rm mess}$ and $M_*$, the operators generating soft scalar mass and the gaugino mass are affected by the hidden sector running.

As an illustrated example, let us take the minimal gauge mediation with the supersymmetry breaking field $\langle X \rangle = M_{\rm mess} + \theta^2 F$.
At the messenger scale $M_{\rm mess}$,
\bea
M_{1/2} & = & \frac{\alpha}{4\pi} \frac{F}{M_{\rm mess}},  \\
m^2 & = & C \left( \frac{\alpha}{4\pi} \right)^2 \left| \frac{F}{M_{\rm mess}} \right|^2. 
\eea
At the scale $M_*$,
\bea
M_{1/2} & = & \left( \frac{M_*}{M_{\rm mess}} \right)^\gamma M_{1/2} (M_{\rm mess}) ,  \\
m^2 & = & \left( \frac{M_*}{M_{\rm mess}} \right)^\Delta m^2 (M_{\rm mess}). 
\eea
Note that the soft scalar mass $m^2$ can be suppressed by
\bea
\left( \frac{M_*}{M_{\rm mess}} \right)^{\Delta-2\gamma},
\eea
compared to $M_{1/2}^2$ and $10^{-3}$ or $10^{-4}$ suppression is easy to obtain with $\Delta \le 1$, $\gamma \simeq 0$ as long as $M_*$ is much lower than $M_{\rm mess}$.
This can provide an effective boundary condition at $M_*$ which resembles the low scale gaugino mediation.

If $M_*$ is low enough (close to the weak scale), we get an effective boundary condition in which all dimension two parameters are much smaller than
the square of the dimension one parameters.
As a result $M_{1/2}, A_t$ and $\mu$ can be larger by a (square root of) loop factor compared to $m$.
Furthermore, if $\mu$ is generated from Giudice-Masiero mechanism (from Kahler potential with the supersymmetry breaking field),
$\mu^2+m_{H_u}^2$ and $\mu^2 + m_{H_d}^2$ appearing in the Higgs mass squared are also suppressed
and large $\mu$ is not directly linked to the electroweak symmetry breaking.
The electroweak symmetry breaking is triggered by the effect of the running below $M_*$ which is mostly given by large $A_t$ term
and is nothing to do with large $\mu$. For $M_*$ close to the weak scale $M_Z$, no large log appears and the electroweak scale
is naturally smaller than the scale of $A_t$ as it appears from the RG running.

The simplest version of the CFT has $F=M_*^2$ as $M_*$ is the only scale existing in the CFT.
In this case the lowest possible $M_*$ is about 1000 TeV when $M_{\rm mess}$ is about $10^4$ TeV to keep $M_{1/2}$ at around 1 TeV.
For large $M_{\rm mess}$, $M_*$ becomes even larger.
However, in principle we can consider $M_*$ close to TeV by allowing a deviation from $F=M_*^2$ in explicit realization of this scenario.
Though it is possible to suppress soft scalar mass enough such that it is much smaller than the gaugino mass,
there is a visible sector loop correction which appears in any case.
Without knowing the detailed knowledge of the strongly coupled hidden sector, we can not compute this effect.
Nevertheless, one loop correction from the visible sector is unavoidable and the small scalar mass squared
is understood up to this one loop threshold correction which is not calculable.
Note that for a certain amount of sequestering the original boundary condition of soft scalar mass is washed out
and is replaced by one loop threshold correction (which is not calculable).

As a result the boundary condition is summarized as follows.
We have large $M_{1/2}$, $A_t$ and $\mu$ while
\bea
m & \simeq & \sqrt{\frac{\alpha}{4\pi}} M_{1/2},
\eea
at $M_*$.
More explicitly,
\bea
m_{\tilde Q}^2 \simeq m_{u^c}^2 \simeq m_{d^c}^2 \simeq m_L^2 \simeq m_{e^c}^2 \ll M_{1/2}^2,  \\
\mu^2 + m_{H_u}^2 \simeq \mu^2 + m_{H_d}^2 \ll M_{1/2}^2, 
\eea
where $\ll$ should be understood as one loop suppression including unknown one loop threshold correction as above.
From now on we do not rely on specific mediation mechanism and will explore the parameter space which keeps the qualitative features
of scalar sequestering (heavy gauginos, heavy higgsinos and possibly large $A_t$ term).

The characteristic features of scalar sequestering is following.
\begin{enumerate}
\item Sfermions (scalars) are lighter than gauginos (fermions).

\item Large $\mu$ does not cause fine tuning in the electroweak symmetry breaking.

\item Higgs fields (scalars) are lighter than higgsinos (fermions).
\end{enumerate}

Before discussing Higgs phenomenology, we just make a short comment on the collider signatures of scalar sequestering.
Below $M_{1/2}$ (symbolically denoting gauginos and higgsinos at the same time), one can write down the effective theory in terms of squarks, sleptons and Higgs in addition to the SM fields
after integrating out gauginos.
Dimension five operators for the sfermions $\phi_i$ and the fermions $\psi_i$ are obtained as following.
\bea
{\cal L} & = & \frac{1}{M_{1/2}} \phi_i^* \phi_j^* \psi_i \psi_j. \nn
\eea
The squarks are produced and have three body decays with a jet, lepton and slepton.
The slepton decays into lepton and goldstino eventually.
The on-shell two body cascade decays do not appear and sharp edge in the invariant mass distribution
will not appear as a consequence.

\section{No Higgs Mixing or Opposite Sign Mixing in the MSSM}

To increase the CP even light Higgs mass, people have concentrated on increasing the quartic of up type Higgs
(for $\tan \beta \ge 3$). This increases the diagonal term and helps raise the light Higgs mass.
However, there is the other way of increasing Higgs mass which is to reduce
the off-diagonal terms in the Higgs mass matrix.

In the MSSM, the tree level mass matrix for the CP even Higgs fields is given by
\bea
{\cal M}^2 & = & \left(
\begin{array}{cc}
M_A^2 \sin^2 \beta + M_Z^2 \cos^2 \beta & - (M_A^2+M_Z^2) \sin \beta \cos \beta \\
- (M_A^2+M_Z^2) \sin \beta \cos \beta & M_Z^2 \sin^2 \beta + M_A^2 \cos^2 \beta
\end{array}
\right), 
\eea
for $(H_d, H_u)$.

For $\tan \beta \ge 3$, it is convenient to write down the expression in terms of $\eta = \frac{1}{\tan \beta}$
and expand it up to ${\cal O}(\eta^2)$. The expression is then
\bea
{\cal M}^2 & = & \left(
\begin{array}{cc}
M_A^2 + (M_Z^2 -M_A^2) \eta^2  & - (M_A^2+M_Z^2) \eta \\
- (M_A^2+M_Z^2) \eta & M_Z^2  + (M_A^2 -M_Z^2) \eta^2
\end{array}
\right). 
\eea
The Higgs mixing angle $\alpha$ is determined from $\tan \beta$ and $M_A$,
\bea
\left( \begin{array}{c} H \\ h \end{array} \right) & = &
\left( \begin{array}{cc} \cos \alpha & \sin \alpha \\
-\sin \alpha & \cos \alpha \end{array} \right)
\left( \begin{array}{c}  {\rm Re } H_d \\  {\rm Re} H_u \end{array} \right), 
\eea
and
\bea
\frac{\tan 2\alpha}{\tan 2\beta} & = & \frac{M_A^2+M_Z^2}{M_A^2-M_Z^2}. 
\eea

Now we include the loop correction to neutral scalar Higgs mass matrix.
We will not consider large $\tan \beta \sim 50$ and bottom Yukawa contribution is negligible.
Then the correction appears as follows.
\bea
{\cal M}^2 & = & \left(
\begin{array}{cc}
M_A^2 + (M_Z^2 -M_A^2) \eta^2 & - (M_A^2+M_Z^2)  \eta + \Delta_{12}\\
- (M_A^2+M_Z^2) \eta +\Delta_{12} & M_Z^2  + (M_A^2 -M_Z^2)  \eta^2 +\Delta_{22}
\end{array}
\right). 
\eea

\bea
\frac{\tan 2\alpha}{\tan 2\beta} & = & \frac{M_A^2+M_Z^2}{M_A^2-M_Z^2 + \frac{\Delta_{22}}{\cos 2\beta}}. 
\eea

When $\Delta_{12}$ is negligible, the eigenvalues are given by
\bea
m_h^2 & = & M_Z^2 (1-4\eta^2) + \Delta_{22},  \\
m_H^2 & = &  M_A^2 + 4 M_Z^2 \eta^2. 
\eea

\begin{figure}[htb]
\includegraphics[width=5.in]{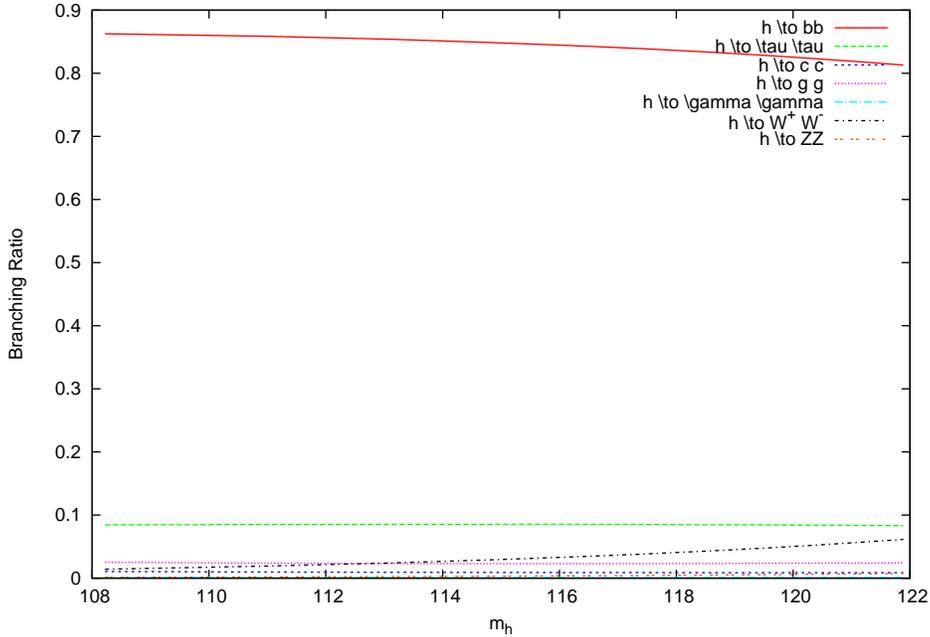}
\caption{MSSM Light Higgs branching ratio plot before including large $\Delta_{12}$
for $\tan \beta=10$ and $M_A=200$ GeV}
\label{fig:MSSMBR}
\end{figure}

Note that an interesting result is obtained if the off-diagonal elements vanishes.
\bea
\Delta_{12} = (M_A^2+M_Z^2) \eta. 
\eea
In this zero mixing angle case, the eigenvalues are read off from the diagonal elements,
\bea
m_h^2 & = & M_Z^2 + (M_A^2 - M_Z^2) \eta^2 + \Delta_{22},  \\
m_H^2 & = & M_A^2 + (M_Z^2 - M_A^2) \eta^2. 
\eea
Compared to the case when the correction to the off-diagonal element is negligible, $\Delta_{12} \sim 0$,
the lightest eigenvalue is increased by $(M_A^2 +3 M_Z^2) \eta^2$
which would be important unless $\eta$ is too small.

No Higgs mixing alters dominant decay channel of Higgs.
For the light Higgs with mass at around 110 GeV to 130 GeV (the MSSM range),
the dominant decay mode is $h \rightarrow b \bar{b}$.

Higgs couplings to the fermions are following.
\bea
h^0 {\bar d} d & : & \lambda_d \frac{\sin \alpha}{\cos \beta},  \\
H^0 {\bar d} d & : & -\lambda_d \frac{\cos \alpha}{\cos \beta}. 
\eea

\begin{figure}[htb]
\subfigure[Light Higgs Branching Ratio]{
\includegraphics[width=3.in]{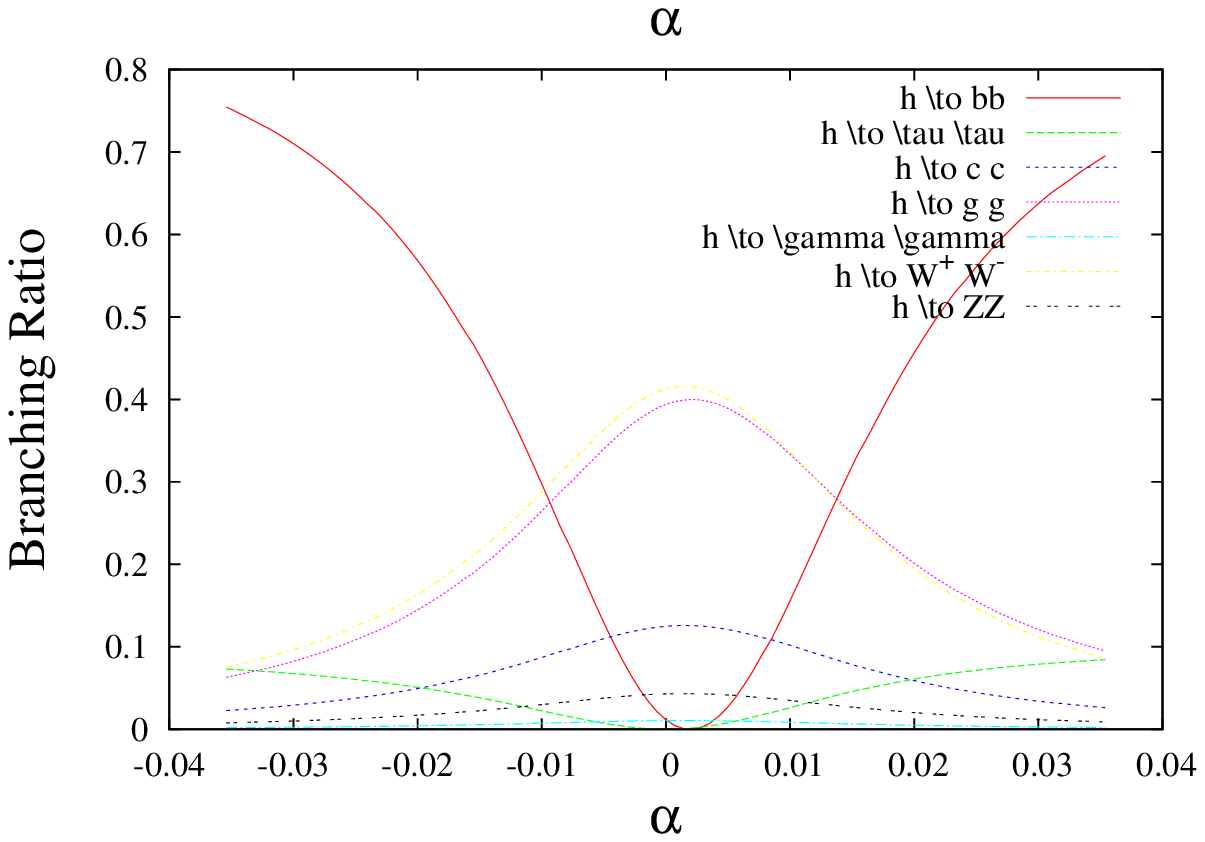}}
\subfigure[Heavy Higgs Branching Ratio]{
\includegraphics[width=3.in]{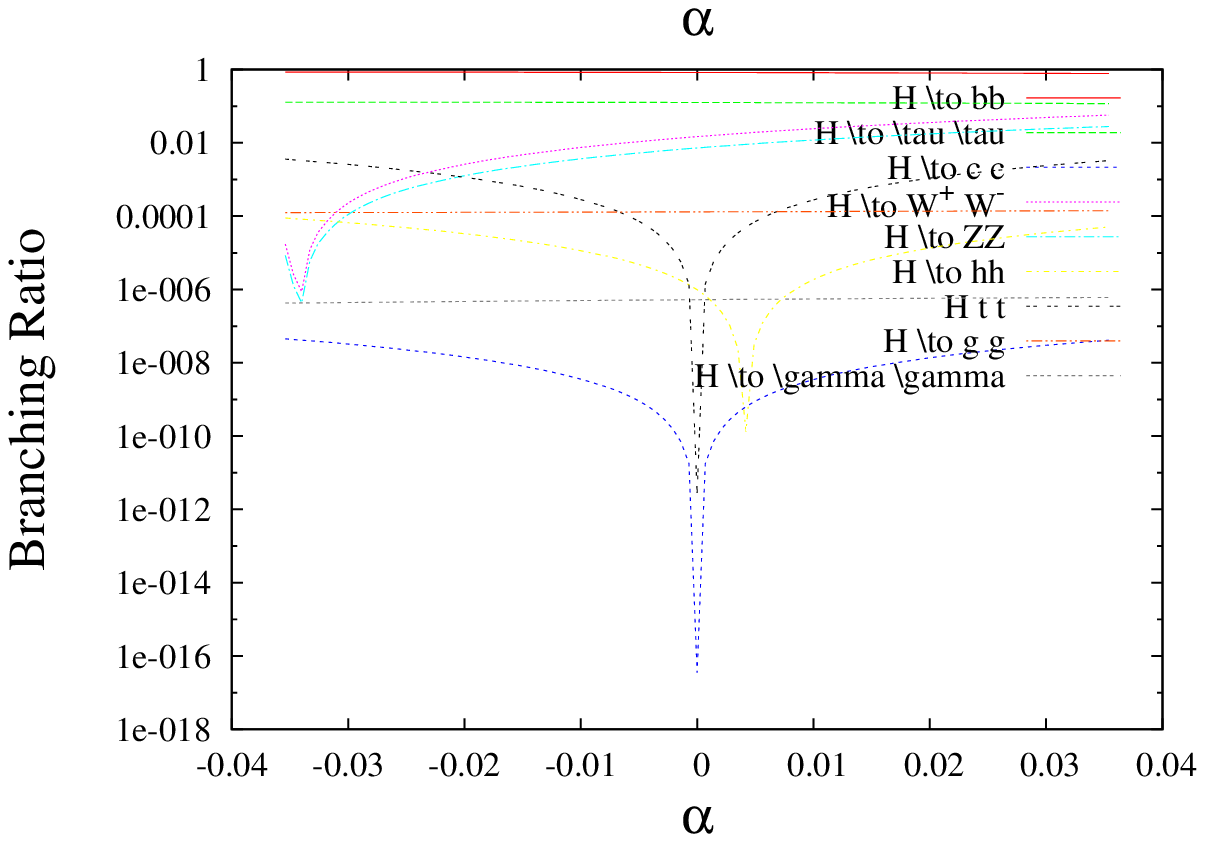}}
\caption{Plot for the branching ratio in terms of mixing angle
for $\tan \beta=10$ and $M_A=200$ GeV}
\label{fig:alphahBR}
\end{figure}

Higgs couplings to the gauge bosons are
\bea
h^0 W^+ W^- & : & g M_W \sin (\beta-\alpha),  \\
H^0 W^+ W^- & : & g M_W \cos (\beta-\alpha), 
\eea
where $M_W$ is the W boson mass.
In the decoupling limit $M_A \rightarrow \infty$, $\beta -\alpha \simeq \frac{\pi}{2}$ and $H^0$ coupling to the gauge boson
is highly suppressed in the usual MSSM.

No Higgs mixing has a direct consequence that the light Higgs is mostly up type and does not couple to $b \bar{b}$.
As a result the partial decay width $\Gamma(h \rightarrow b \bar{b})$ is suppressed
while other decay widths are not changed much.
Hence, the branching ratio $B(h\rightarrow W^{+*} W^{_*}) = \Gamma (h\rightarrow W^{+*} W^{_*})/ \Gamma(h\rightarrow {\rm all})
\simeq \Gamma (h\rightarrow W^{+*} W^{_*})/ \Gamma(h\rightarrow b \bar{b})$ can be highly enhanced. 
The associated production with a vector boson $(q \bar{q^{\prime}} \rightarrow W h)$ can give a trilepton signature
which would make it possible to study the gauge interactions of the light Higgs \cite{Baer:1998cm}.
The same is true for the $B(h \rightarrow \gamma \gamma)$ enhancement.
($\Gamma (h \rightarrow W^{+*} W^{-*})$ and $\Gamma (h \rightarrow \gamma \gamma)$ have little change.)

For the heavy Higgs, the coupling to the gauge boson is increased from $\cos (\beta -\alpha ) \sim 0$ to $\cos \beta \simeq \eta$.
Then the production cross section of the heavy Higgs is enhanced and makes it easier to access the heavy Higgs at the LHC
relatively compared to the case in which $\cos (\beta-\alpha) \sim 0$ is maintained.

We consider two most important corrections to the Higgs effective potential in the following two sections.
Section A discusses the correction to the Higgs mass
due to the change of $(H_u^\dagger H_u)^2$ term.
Section B deals with the change of $H_u H_d (H_u^\dagger H_u)$ term which contributes
to the mass and modifies Higgs decay.

\subsection{$(H_u^\dagger H_u)^2$}

\begin{center}
\begin{picture}(310,110)(-10,-10)
\DashLine(0,0)(100,80){5}
\DashLine(100,0)(0,80){5}
\DashLine(70,24)(70,56){5}
\Text(0,80)[r]{$H_u^*$}
\Text(0,0)[r]{$H_u$}
\Text(100,0)[t]{$H_u^*$}
\Text(100,80)[b]{$H_u$}
\Text(70,56)[b]{$A_t$}
\Text(70,24)[t]{$A_t$}
\Text(150,80)[r]{$H_u^*$}
\Text(150,0)[r]{$H_u$}
\Text(250,0)[t]{$H_u^*$}
\Text(250,80)[b]{$H_u$}
\Text(175,60)[r]{$A_t$}
\Text(225,60)[b]{$A_t$}
\Text(175,20)[r]{$A_t$}
\Text(225,20)[t]{$A_t$}
\DashLine(150,80)(175,60){5}
\DashLine(150,0)(175,20){5}
\DashLine(250,0)(225,20){5}
\DashLine(250,80)(225,60){5}
\DashLine(175,60)(225,60){5}
\DashLine(175,60)(175,20){5}
\DashLine(175,20)(225,20){5}
\DashLine(225,20)(225,60){5}
\end{picture}
\end{center}

Let us consider the correction to the Higgs mass from the diagonal element.
One loop logarithmic correction between stop mass and top mass is well known to give a large correction to the Higgs mass.
However, 114 GeV bound from LEP II can be achieved only for sizable separation of stop mass and top mass, e.g., $m_{\tilde{t}} =1$ TeV,
and this threaten the naturalness of weak scale supersymmetry as we typically need a percent or worse fine tuning to get $M_Z$ correctly.
The fine tuning can be ameliorated if the cutoff (more precisely the scale at which the running of soft parameters start, for instance,
messenger scale in gauge mediation) is low enough but still it is a few percent.
Thus we focus on the case in which stop is not so heavy, $m_{\tilde t} \le 500$ GeV. Then there are other corrections which are as important as top stop loop
or even more important which are the finite threshold corrections obtained when stops are integrated out.
They provide effective dimension six and eight operators suppressed by stop mass squared and square of it.
Soft tri-linear term generates
\bea
V(H_u, H_d) & = & \epsilon_2 (H_u^* H_u)^2 , 
\eea
with a coefficient
\bea
\epsilon_2 & = & \frac{3 y_t^4}{16\pi^2} \frac{A_t^2}{m_{\tilde{t}}^2} \left[ 1 - \frac{A_t^2}{12 m_{\tilde{t}}^2} \right], 
\eea
where $y_t$ is the top Yukawa coupling and $m_{\tilde t}^2 = (m_{\tilde Q_3}^2 + m_{\tilde t^c}^2)/2$ is used.
This correction is maximized when $A_t = \pm \sqrt{6} m_{\tilde{t}}$ which gives
\bea
\epsilon_2 ({\rm max}) & = & \frac{9y_t^4}{16\pi^2}. 
\eea
To discuss the Higgs mass in terms of Higgs Vacuum Expectation Value(VEV), we use the convention $M_Z^2 = \frac{1}{2}(g^2+g^{\prime 2}) (v_u^2+v_d^2) = \frac{1}{2}(g^2+g^{\prime 2}) v^2$.
The correction to the Higgs mass (when $\eta \ll 1$) is then
\bea
\delta_{\epsilon_2} {\cal M}^2 & = & \left( \begin{array}{cc}
0 & 0 \\
0 & 4\epsilon_2 ({\rm max}) v^2 \end{array} \right) \simeq
\left( \begin{array}{cc}
0 & 0 \\
0 &  0.23 v^2 \end{array} \right)
 . 
\eea
This correction alone is large enough to increase Higgs mass from 90 GeV to 120 GeV.
The same correction can be obtained from the logarithmic correction if $\log \frac{m_{\tilde t}}{m_t} \simeq 3$, i.e., $m_{\tilde t} \simeq 2$ TeV
which causes a serious fine tuning in the electroweak symmetry breaking.

\subsection{$H_u H_d (H_u^\dagger H_u)$}

\begin{center}
\begin{picture}(310,110)(-10,-10)
\DashLine(0,0)(100,80){5}
\DashLine(100,0)(0,80){5}
\DashLine(70,24)(70,56){5}
\Text(0,80)[r]{$H_u^*$}
\Text(0,0)[r]{$H_u$}
\Text(100,0)[t]{$H_d$}
\Text(100,80)[b]{$H_u$}
\Text(70,56)[b]{$A_t$}
\Text(70,24)[t]{$\mu$}
\Text(150,80)[r]{$H_d$}
\Text(150,0)[r]{$H_u$}
\Text(250,0)[t]{$H_d$}
\Text(250,80)[b]{$H_u$}
\Text(175,60)[r]{$A_t$}
\Text(225,60)[b]{$A_t$}
\Text(175,20)[r]{$A_t$}
\Text(225,20)[t]{$\mu$}
\DashLine(150,80)(175,60){5}
\DashLine(150,0)(175,20){5}
\DashLine(250,0)(225,20){5}
\DashLine(250,80)(225,60){5}
\DashLine(175,60)(225,60){5}
\DashLine(175,60)(175,20){5}
\DashLine(175,20)(225,20){5}
\DashLine(225,20)(225,60){5}
\end{picture}
\end{center}

\begin{figure}[htb]
\subfigure[$\alpha$]{
\includegraphics[width=3.in]{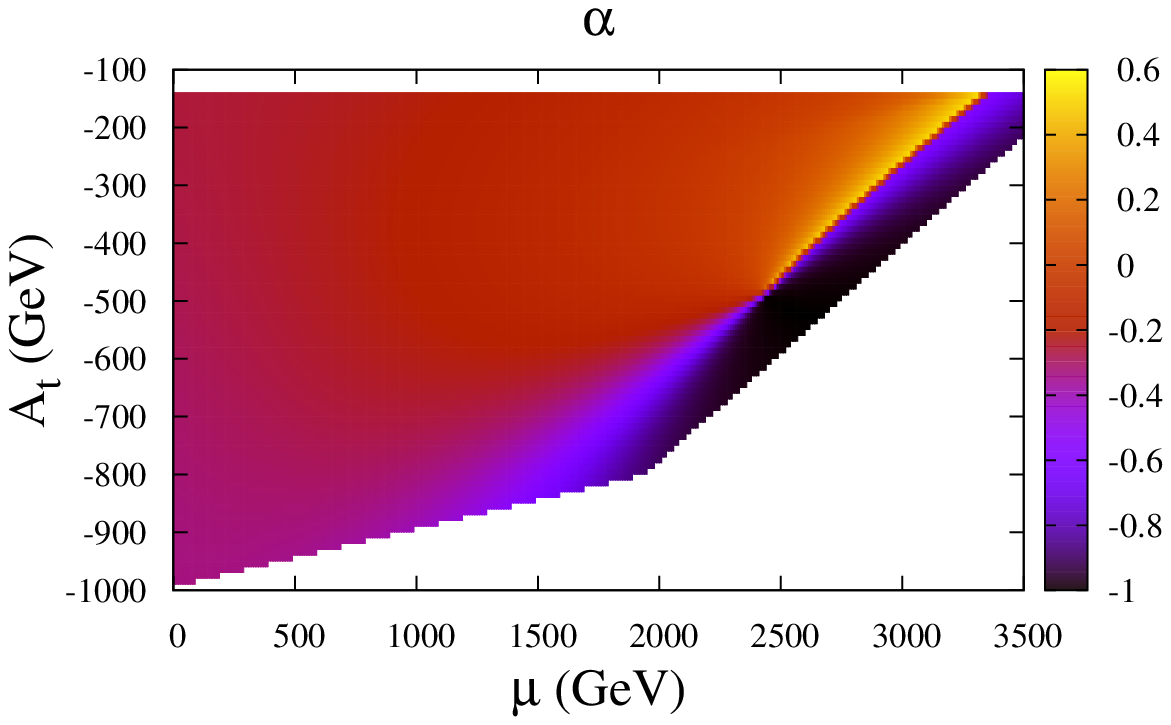}}
\subfigure[$B(h\rightarrow b\bar{b})$]{
\includegraphics[width=3.in]{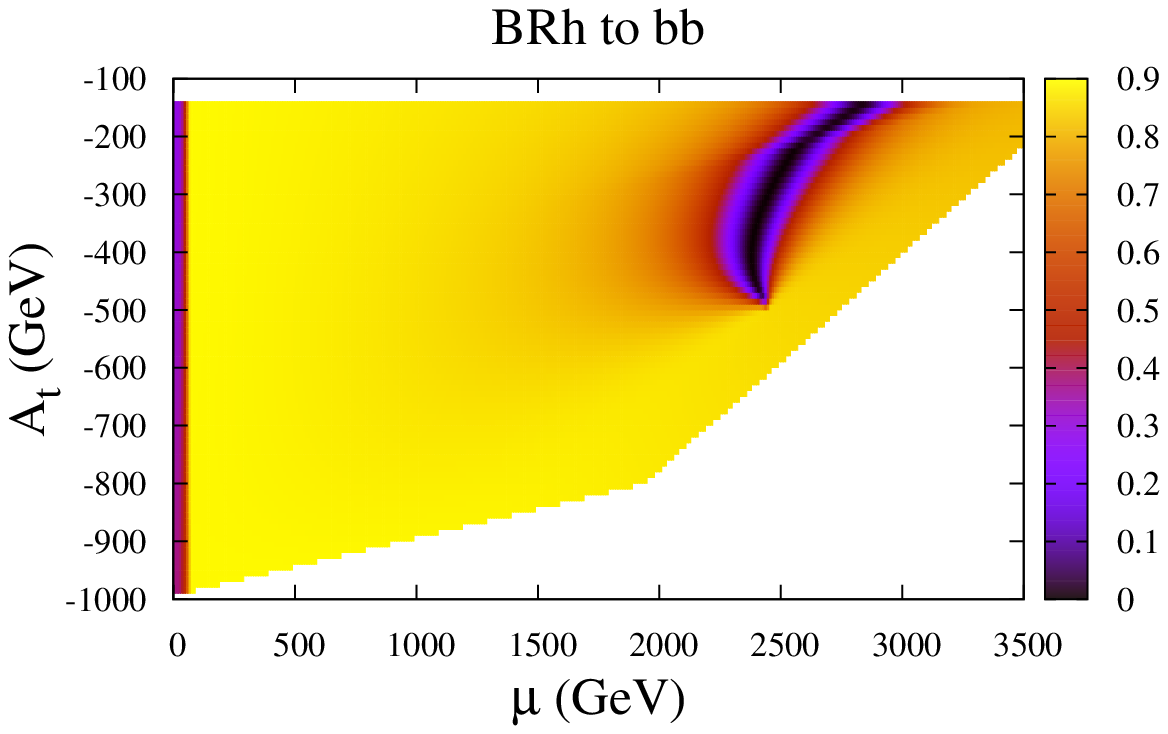}}
\caption{Plot of $\alpha$ and $B(h\rightarrow b\bar{b})$ with $A_t$ and $\mu$ scan for $\tan \beta=10$, $M_A=150$ GeV and $M_{\tilde t}=500$ GeV}
\label{fig:muAtmixingangle}
\end{figure}

Similarly we can obtain
\bea
V(H_u, H_d) & = & -\epsilon_1 H_u H_d (H_u^* H_u) + {\rm h.c.}= -\epsilon_1 (H_u^+ H_d^- - H_u^0 H_d^0) (H_u^{+*} H_u^+ + H_u^{0*} H_u^0) +{\rm h.c.}, 
\eea
by replacing one vertex to $\mu$ rather than $A_t$ (such that $H_u^*$ is replaced by $H_d$).
\bea
\epsilon_1 & = & -\frac{3y_t^4}{16\pi^2}  \frac{\mu A_t}{m_{\tilde{t}}^2} \left[ 1 - \frac{A_t^2}{6 m_{\tilde{t}}^2} \right],
\nn
\eea
up to ${\cal O}(\eta)$.

The correction to the Higgs mass matrix appears in ${\cal M}_{11}^2$ and ${\cal M}_{12}^2$ (and $21$).
\bea
\delta_{\epsilon_1} {\cal M}_{22}^2 & = & 4\epsilon_1 v_u^2 \eta = -\frac{3y_t^4}{4\pi^2}  \frac{\mu A_t }{m_{\tilde{t}}^2} \left[ 1-\frac{A_t^2}{6 m_{\tilde{t}}^2} \right] v^2 \eta
+ {\cal O}(\eta^2), 
\eea
and
\bea
\delta_{\epsilon_1} {\cal M}_{12}^2 & = & 2\epsilon_1 v_u^2 = -\frac{3y_t^4}{8\pi^2} \frac{\mu A_t }{m_{\tilde{t}}^2} \left[ 1- \frac{A_t^2}{6 m_{\tilde{t}}^2} \right] v^2 + {\cal O}(\eta), 
\eea
which can have a maximum contribution when
\bea
A_t & = & -\sqrt{2} m_{\tilde{t}}, 
\eea
and the correction is then
\bea
\epsilon_1 ({\rm max}) & = & \frac{\sqrt{2} y_t^4}{8\pi^2} \frac{\mu}{m_{\tilde{t}}}. 
\eea
The correction appears in the Higgs mass matrix as follows (up to ${\cal O}(\eta)$ and ${\cal O}(\eta^2)$ respectively).
\bea
\delta_{\epsilon_1} {\cal M}^2 & = & \left( \begin{array}{cc}
0 & 2\epsilon_1  \\
2\epsilon_1  & 4\epsilon_1 \eta
\end{array} \right) v^2 . 
\eea

\begin{figure}[htb]
\subfigure[$m_h$]{
\includegraphics[width=3.in]{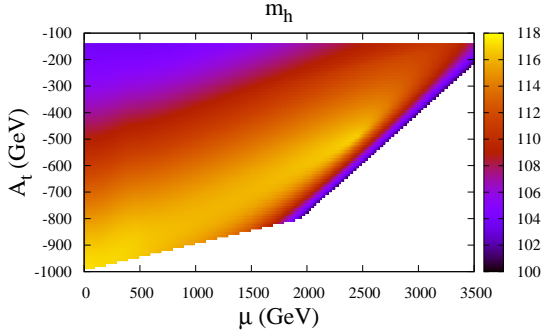}}
\subfigure[$\mu$ vs $m_h$ for $A_t = -500$ GeV]{
\includegraphics[width=2.5in]{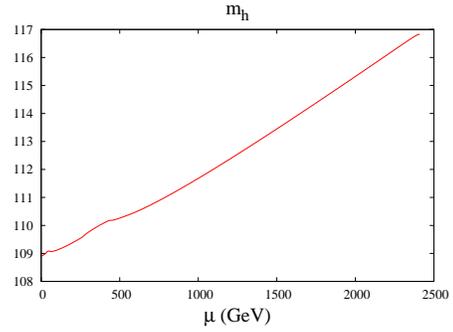}}
\subfigure[Threshold correction/$|$Tree$|$ in 12 element]{
\includegraphics[width=3.in]{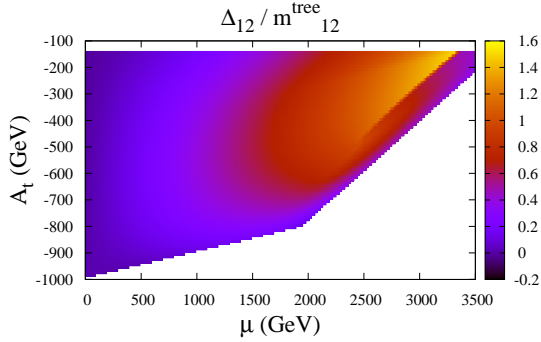}}
\subfigure[Threshold correction/Tree in 22 element]{
\includegraphics[width=3.in]{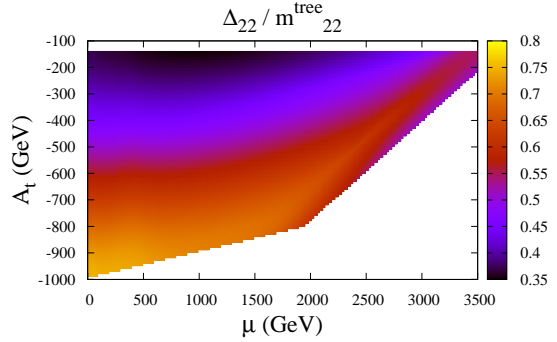}}
\caption{Plot of $m_h$ and each correction to the Higgs mass matrix with $A_t$ and $\mu$ scan for $\tan \beta=10$, $M_A=150$ GeV and $M_{\tilde t}=500$ GeV}
\label{fig:muAtmh}
\end{figure}

It is possible to have a sizable $\epsilon_1$( $\epsilon_1/|{\cal M}_{12 \ {\rm tree}}^2| \sim 1$ or $2$) if $\mu$ is large compared to the stop mass $m_{\tilde{t}}$.
$\mu$ can not be arbitrarily large as the next order correction which we neglect here comes with ${\cal O} \mu^2/m_{\tilde t}^2 \eta$ which may be as important as $\mu/m_{\tilde t}$ depending on $\eta$.
The contribution to the diagonal entry is suppressed by $\tan \beta$ and does not have a significant role
for $\tan \beta \ge 5$ or $10$.
It is the off-diagonal element which can contribute significantly to the light Higgs mass.
Increasing $\mu$ changes mixing angle and the branching ratio of Higgs decay significantly.
$\mu$ can be very large in scalar sequestering scenario
without causing fine tuning problem in the electroweak symmetry breaking
as $\mu^2 + m_{H_u}^2$ and $\mu^2+m_{H_d}^2$ are suppressed to be close to zero.
In other words, the conventional range of $\mu \le$ 1 TeV
is enlarged to $\mu \le \sqrt{\frac{2\pi}{3\alpha \log (M_*/M_Z)}} {\rm TeV} \sim 5$ TeV
($\log M_*/M_Z \simeq 5$ is taken).
No mixing can happen even for $\tan \beta =10$ or smaller.

There is an interesting regime in addition to no Higgs mixing regime in which the understanding of the effects coming from the one loop correction
generating $H_u H_d H_u^* H_u$ is important.
It is when the mixing angle $\alpha$ flips its sign due to the loop correction.
It can happen when $\Delta_{12} \sim 2 (M_A^2+M_Z^2) \eta$.
The contribution of the off-diagonal element to the eigenvalue is similar (or slightly larger) compared to the case $\Delta_{12}=0$.
Now the contribution to the diagonal element can increase the light Higgs mass which is about $4\epsilon_1 \eta \simeq 4(M_A^2+M_Z^2) \eta^2$.
This correction can be very large when $\eta \le 1/10$ or $1/5$.

Indeed we can plot the lightest Higgs mass as a function of $\epsilon_1$ and the maximum arises before $\epsilon_1$ becomes too large compared
to ${\cal M}_{12}$. The exact value of $\epsilon_1$ which gives the maximum Higgs mass is a function of $M_A$ and $\eta$.

\section{No Higgs Mixing in the BMSSM}

In this section the same physics in the beyond the MSSM(BMSSM) is briefly discussed.
In \cite{Dine:2007xi}, it is claimed that sizable correction for the light Higgs mass is possible
in case when there is a new particle coupling to the Higgs at around TeV.
Sizable correction at $\tan \beta=10$ is a rather surprising result considering the NMSSM in which the new coupling brings a new quartic
to the Higgs doublets but only with a suppression of $\eta^2$.
Furthermore there is a factor 2 discrepancy between \cite{Dine:2007xi} and the earlier work \cite{Brignole:2003cm}.

After careful examination of the BMSSM analysis, the factor 2 difference is understood as follows.
In \cite{Brignole:2003cm}, the correction to the light Higgs mass has been computed by considering the new quartic term of the effective operator.
In \cite{Dine:2007xi}, the full correction to the light Higgs mass is obtained since the contribution of the new operator to the off-diagonal element
is also taken into account. Note that the light Higgs mass becomes lighter after diagonalization since the off-diagonal element makes the light one
to be ligher (and the heavy one to be heavier). If the new operator appears in the off-diagonal element to cancel the effect
of the original off-diagonal element, it can help increase the light Higgs mass.
This brings a factor two discrepancy between the results given in two papers.
As a result of reducing off-diagonal element, the validity of the eq. (31) in \cite{Dine:2007xi} is limited to the case
in which the correction of the dimension five operator in the off-diagonal element is smaller than the original off-diagonal element.
If the correction of the dimension five operator is larger than the original off-diagonal element,
it makes the light eigenvalue to be lighter as the total off-diagonal element becomes larger as we increase the new correction.
The tree level (or MSSM) off-diagonal element is $-(M_A^2+M_Z^2) \eta$ and, for a given $M_A$ and $\eta$, there is a hidden constraint which shows when the eq. (31) in \cite{Dine:2007xi} breaks down.

Let us discuss the physics slightly more detail.
The pseudoscalar mass is given in terms of the following two by two matrix. $\epsilon$ here corresponds to $\epsilon_{1r}$ in \cite{Dine:2007xi}.
\bea
{\cal M}_{\rm Im}^2 & = & (B\mu - 2 \epsilon v^2) \left(
\begin{array}{cc}
\frac{v_d}{v_u} & 1 \\
1 & \frac{v_u}{v_d}
\end{array}
\right).
\eea
After diagonalization, we obtain
\bea
M_{G_0}^2 & = & 0,   \\
M_A^2 & = & (B\mu - 2\epsilon v^2)\eta. 
\eea

\begin{figure}[thb]
\subfigure[$\epsilon$ vs Higgs mass]{
\includegraphics[width=3.in]{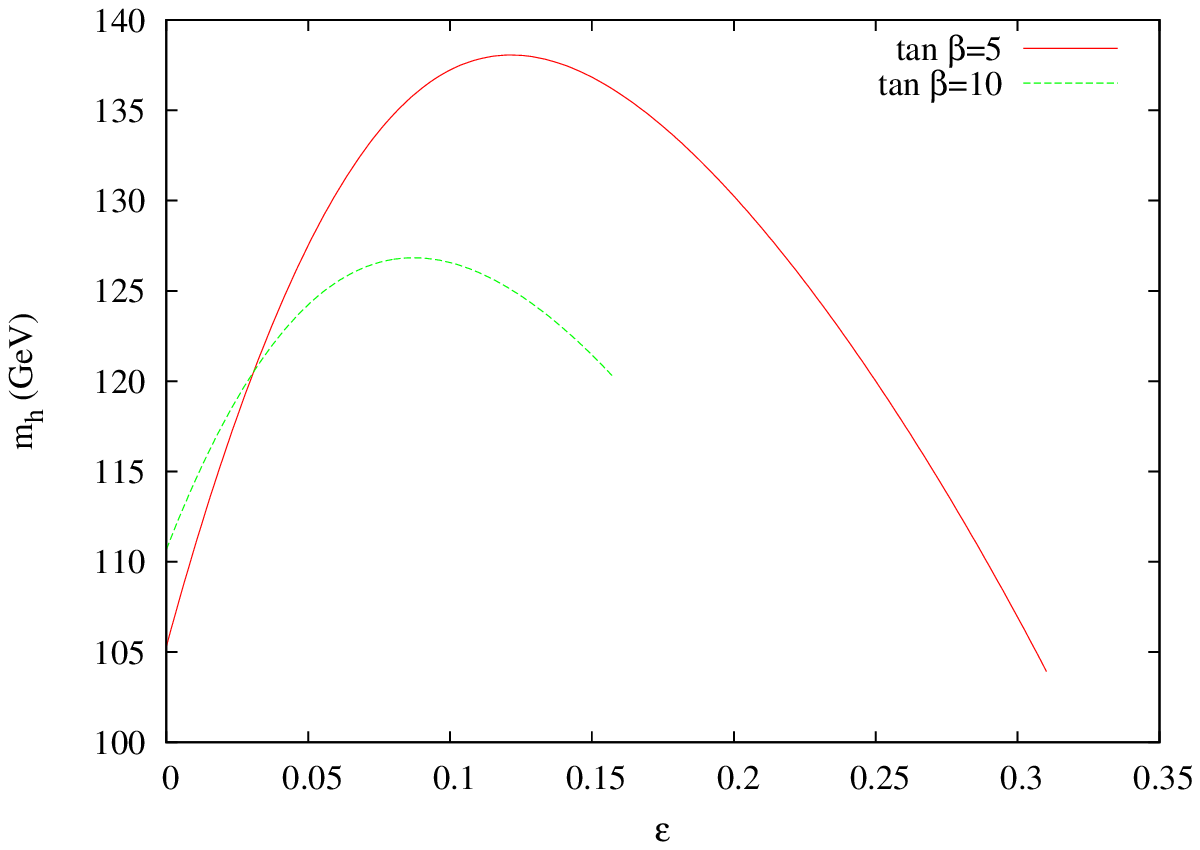}}
\subfigure[$\epsilon$ vs mixing angle]{
\includegraphics[width=3.in]{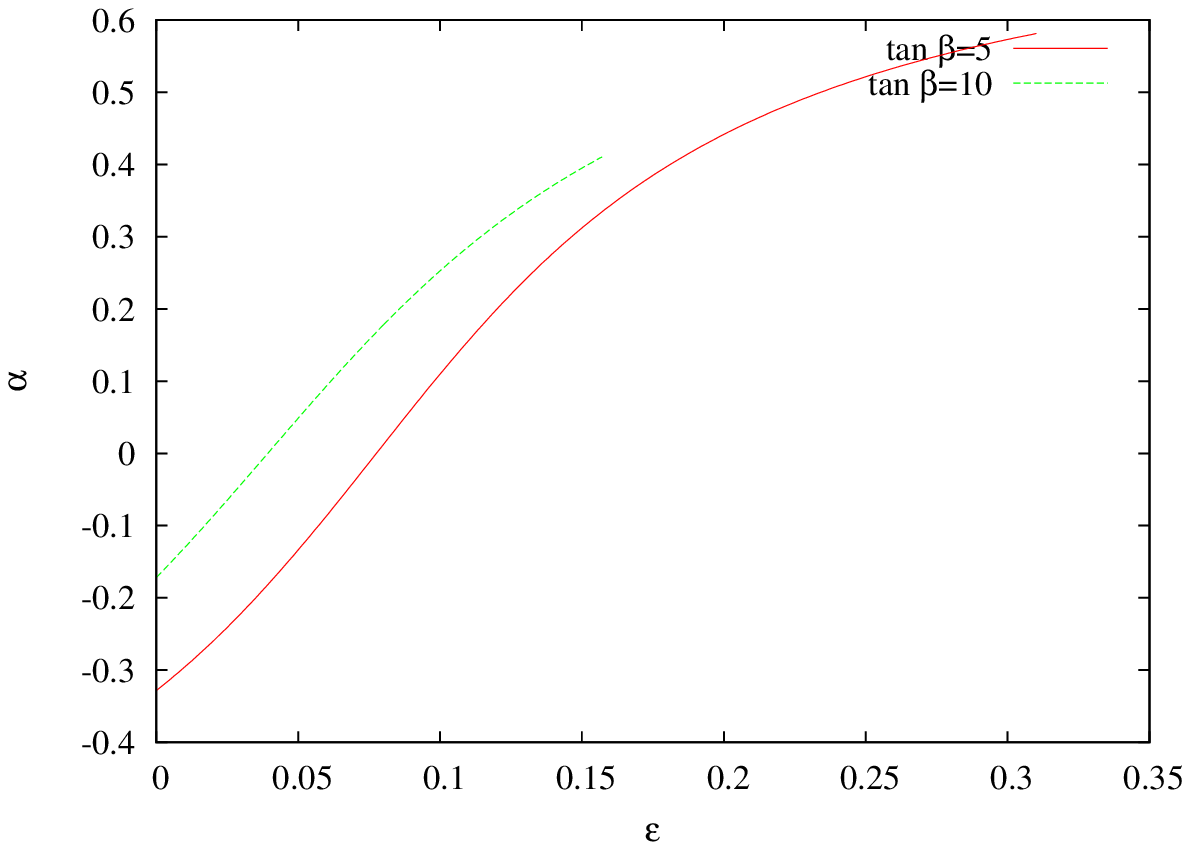}}
\caption{Plot of $\epsilon$ vs $m_h$ and $\alpha$ for $\tan \beta=10$ and $M_A = 300$ GeV}
\label{fig:BMSSM}
\end{figure}

The effect of $\epsilon$ is absorbed into the relation of $M_A$ and $B\mu$.
For the scalar Higgs mass matrix, the full expression up to $\eta^2$ and $\epsilon \eta$ is
\bea
{\cal M}_{\rm Re}^2 & = & \left( \begin{array}{cc}
M_A^2  + (M_Z^2-M_A^2) \eta^2  + 8\epsilon v^2 \eta  & -(M_A^2+Z_Z^2) \eta + 4 \epsilon v^2 \\
-(M_A^2+M_Z^2) \eta + 4 \epsilon v^2 & M_Z^2 + (M_A^2-M_Z^2) \eta^2 + 8\epsilon v^2 \eta + \Delta_{22}
\end{array}
\right), 
\eea
where $\Delta$ is the loop correction mainly coming from top loop but we neglect it in the following discussion.
With the inclusion of $\epsilon$ correction, the diagonal one gets a correction $8\epsilon v^2 \eta$ for both $m_h^2$ and $m_H^2$
and the off-diagonal one gets $4\epsilon v^2$.

We can consider three different regimes depending on the relative size of $\epsilon$.
\begin{itemize}

\item {Small correction regime, $4\epsilon v^2 \ll (M_A^2+M_Z^2) \eta$}

\end{itemize}

The expression in \cite{Dine:2007xi} is valid.
However, at the same time, you get a constraint on the largest possible size of $\epsilon$.
If $4\epsilon v^2 \ll (M_A^2+M_Z^2) \eta$,
the largest possible correction to the light Higgs mass is bounded by
\bea
16 \epsilon v^2 \eta & \ll & 4 (M_A^2+M_Z^2) \eta^2. 
\eea
For $\eta \simeq 0.1$, the correction is very small unless $M_A$ is large enough.
The inequality limits the possible correction to the light Higgs mass.
For $M_A = 300$ GeV and $\tan \beta = 10$, the largest possible $\epsilon$ which keeps the validity of the expression is when
$16 \epsilon v^2 \simeq (M_A^2+M_Z^2) \eta^2$ and it increases the Higgs mass just by 5 GeV (from 90 GeV to 95 GeV).

\begin{itemize}

\item {No mixing regime, $4\epsilon v^2 \sim (M_A^2+M_Z^2) \eta$}

\end{itemize}

No mixing regime is very interesting as it changes the branching fraction of Higgs entirely.
With a cancellation of the off-diagonal element, the decay to $b \bar{b}$ is suppressed and the couplings of the Higgs to down type quarks and charged leptons
are highly reduced.
Hence, other decay modes can have a sizable branching fraction, e.g., $h \rightarrow W W^*$.
(The partial decay width $\Gamma (h \rightarrow W W^*)$ is hardly changed.)
For the Higgs mass, the prediction is very easy.
For given $m_A$ and $\tan \beta$,
we can fix the $\epsilon$ correction from the no mixing relation.
\bea
4 \epsilon v^2 & = & (M_A^2+M_Z^2) \eta. 
\eea
Now the eigenvalue is just the diagonal element.
\bea
m_h^2 & = &  M_Z^2 + (M_A^2-M_Z^2) \eta^2 + 8 \epsilon v^2 \eta = M_Z^2 + (3M_A^2+M_Z^2) \eta^2. 
\eea
For $\tan \beta =10$, $M_A=300$ GeV, we can increase $m_h = 90$ GeV to $m_h = 105$ GeV.

\begin{itemize}

\item {Large correction regime, $4\epsilon v^2 \gg (M_A^2+M_Z^2) \eta$}

\end{itemize}

The maximum of the light Higgs mass occurs roughly when the off-diagonal elements flip the sign by $\epsilon$ correction
though the exact condition for the maximum depends on whole matrix elements ($M_A$ and $\eta$).
\bea
4\epsilon v^2 & \sim & 2 (M_A^2+M_Z^2) \eta. 
\eea
If the $\epsilon$ correction is larger than this, the off-diagonal elements will reduce the light eigenvalue in the diagonalization
which overcome the gain coming from the diagonal term.
\bea
m_h^2 & = &   M_Z^2 + (5M_A^2+3M_Z^2) \eta^2. 
\eea
For the same sample point, $\tan \beta =10$ and $M_A = 300$ GeV,
we can increase $m_h$ from 90 GeV to 114 GeV.
Of course it is important to consider the higher order correction like $\epsilon^2$ and other higher dimensional operators.
It is important that the Higgs mixing angle $\alpha$ between $h$ and $H$ has the opposite sign
compared to the usual MSSM.
The change of the sign affects the branching ratio from the interference terms.
Even for the small mixing angle $\sin \alpha \sim 0.1$, the partial decay width $\Gamma (H \rightarrow W^+ W^-)$ can change from -33 \% to 50 \%.

\section{Conclusion}

No discovery of supersymmetric particles puts a strong constraint on the weak scale supersymmetry.
Most of all the MSSM is seriously threatened by the Higgs mass bound. To understand why the scale of the electroweak symmetry breaking
is so low compared to other sparticle masses would be the main issue concerning the weak scale supersymmetry (especially the MSSM).
In this paper we use this constraint to understand the sparticle spectrum of the MSSM.
Rather than relying on the usual logarithmic correction from the top loop which causes a serious fine tuning problem,
we focus on finite threshold corrections arising when stops are integrated out.
Maximal stop mixing scenario is one of the most attractive possibilities from this point of view.

In addition, the idea of scalar sequestering using the strongly coupled hidden sector brings an entirely new patterns of sparticle spectrum.
Sfermions and Higgs fields are light while gauginos are higgsinos are heavy. If `generalized' Giudice-Masiero mechanism is responsible
for the generation of the $\mu$ term, the electroweak scale can be smaller by (the square root of ) the loop factor compared to $\mu$ (the higgsino mass).
Naturally large $\mu$ opens new possibility of increasing the Higgs mass by cancelling the off-diagonal elements using the threshold correction easily by 10 to 20 GeV. This correction at the same time alters the Higgs mixing angle entirely
and the Higgs decay branching fraction is drastically modified as a consequence. The same phenomenology happens even for small $\mu$ if $\tan \beta$ is large enough
as the off-diagonal element is negligible from the beginning. In this case, however, the correction to the Higgs mass is very small and the correction just alters the Higgs mixing angle.

For large $\mu$ and moderate $\tan \beta$ as discussed in this paper, the same correction not only changes the Higgs mixing angle but also increases the light Higgs mass significantly. As a result the LEP bound on the light Higgs mass can be explained without heavy stop mass
and leaves a room for a natural understanding of the electroweak symmetry breaking in the MSSM. The suppression of $h \rightarrow b\bar{b}$ branching ratio and the enhancement of $h \rightarrow W^{+*} W^{-*}$ branching ratio as a result might allow unexpected early discovery of the light Higgs using the trilepton signature at the LHC. The heavy Higgs also has an enhanced production cross section as the coupling to the gauge boson is enhanced compared to the case of no sizable $\Delta_{12}$ correction.
We leave a detailed study of discovery potential of the Higgs at the LHC with this scenario as a future work.

\section{Acknowledgement}

We thank N. Arkani-Hamed, D. O'Connell, R. Rattazzi and N. Seiberg for discussions.
JK thanks to the Institute for Advanced Study at Princeton for the hospitality during his visit while this work has been done.
This work is supported by KRF-2008-313-C00162 and CQUeST of Sogang University.

\end{document}